\documentclass[11pt]{article}
\usepackage{graphicx}
\usepackage{authblk}
\usepackage[margin=1.25in]{geometry}
\usepackage[usenames,dvipsnames]{color}
\usepackage{url}
\usepackage[colorlinks = true,
            linkcolor = blue,
            urlcolor  = blue,
            citecolor = blue,
            anchorcolor = blue]{hyperref}

\newcommand\pubnumber{Transcendental Preprint }
\newcommand\pubdate{\today}

\def\Title#1{\begin{center} {\LARGE #1 } \end{center}}
\def\Author#1{\begin{center}{ \sc #1} \end{center}}

\newcommand\pubblock{\rightline{\begin{tabular}{l} \pubnumber\\
         \pubdate \end{tabular}}}
\newenvironment{Abstract}{\begin{quotation} \begin{center}
                       ABSTRACT
     \end{center}\bigskip  }{\end{quotation}}

\def\beq{\begin{equation}}
\def\eeq#1{\label{#1}\end{equation}}
\def\eeqn{\end{equation}}

\newenvironment{Eqnarray}%
   {\arraycolsep 0.14em\begin{eqnarray}}{\end{eqnarray}}
\def\beqa{\begin{Eqnarray}}
\def\eeqa#1{\label{#1}\end{Eqnarray}}
\def\eeqan{\end{Eqnarray}}

\let\bar=\overbar

\def\lsim{\mathrel{\raise.3ex\hbox{$<$\kern-.75em\lower1ex\hbox{$\sim$}}}}
\def\gsim{\mathrel{\raise.3ex\hbox{$>$\kern-.75em\lower1ex\hbox{$\sim$}}}}

\def\del{\partial}
\def\Dslash{\not{\hbox{\kern-4pt $D$}}}
\def\dslash{\not{\hbox{\kern-2pt $\del$}}}
\def\pslash{\not{\hbox{\kern-2pt $p$}}}
\def\ETmiss{\not{\hbox{\kern-4pt $E$}}_T}

\def\Dlr{\mathrel{\raise1.5ex\hbox{$\leftrightarrow$\kern-1em\lower1.5ex\hbox{$D$}}}}

\def\MSB{{\bar{M \kern -2pt S}}}
\def\msb{{\bar{\scriptsize M \kern -1pt S}}}

\def\drb{{\bar{\scriptsize D \kern -1pt R}}}

\newcommand\snowmass{\begin{center}\rule[-0.2in]{\hsize}{0.01in}\\\rule{\hsize}{0.01in}\\
\vskip 0.1in Submitted to the  Proceedings of the US Community Study\\ 
on the Future of Particle Physics (Snowmass 2021)\\ 
\rule{\hsize}{0.01in}\\\rule[+0.2in]{\hsize}{0.01in} \end{center}}

\begin{document}

\pubblock

\Title{Smart sensors using artificial intelligence \\ for on-detector electronics and ASICs}

\small
\Author{
	\textbf{Gabriella Carini, Grzegorz Deptuch, Jin Huang, Soumyajit Mandal, Sandeep Miryala, Veljko Radeka, Yihui Ren} \\
	Brookhaven National Laboratory \\
	\textbf{Jennet Dickinson, Farah Fahim, Christian Herwig, Cristina Mantilla Suarez, Benjamin Parpillon, Nhan Tran} \\
	Fermi National Accelerator Laboratory  \\
	\textbf{Philip Harris, Dylan Rankin} \\
	Massachusetts Institute of Technology  \\
	\textbf{Dionisio Doering, Angelo Dragone, Ryan Herbst, Lorenzo Rota, Larry Ruckman} \\
	SLAC National Accelerator Laboratory  \\
	\textbf{Allison McCarn Deiana} \\
	Southern Methodist University  \\
	\textbf{F. Mitchell Newcomer} \\
	University of Pennsylvania  \\
}
\normalsize


\begin{Abstract}
\noindent Cutting edge detectors push sensing technology by further improving spatial and temporal resolution, increasing detector area and volume, and generally reducing backgrounds and noise.  This has led to a explosion of more and more data being generated in next-generation experiments.  Therefore, the need for near-sensor, at the data source, processing with more powerful algorithms is becoming increasingly important to more efficiently capture the right experimental data, reduce downstream system complexity, and enable faster and lower-power feedback loops.  In this paper, we discuss the motivations and potential applications for on-detector AI.  Furthermore, the unique requirements of particle physics can uniquely drive the development of novel AI hardware and design tools. We describe existing modern work for particle physics in this area.  Finally, we outline a number of areas of opportunity where we can advance machine learning techniques, codesign workflows, and future microelectronics technologies which will accelerate design, performance, and implementations for next generation experiments.
\end{Abstract}

\snowmass
\def\thefootnote{\fnsymbol{footnote}}
\setcounter{footnote}{0}

\pagebreak
\tableofcontents

\pagebreak



\section{Executive Summary}
Modern particle physics experiments and accelerators, exploring nature at increasingly finer spatial and temporal scales in extreme environments, create massive amounts of data which require real-time data reduction as close to the data source and sensors as possible.  
Due to historical requirements for processing data in extreme environments in high energy physics experiments, the advancement of AI in near-detector electronics is a natural capability for detector readout in the field and can be a driver for progress in other scientific domains.
This will enable more powerful data compression and filtering which better preserves the physics content of experiments, reduces downstream system complexity, and provides fast feedback and control loops.  

While the emergence of AI is growing rapidly across science and industry, the requirements of particle physics for speed, throughput, fidelity, interpretability, and reliability in extreme environments requires advancing state-of-the-art technology in use-cases that go far beyond industrial and commercial applications.  Therefore, dedicated investment in powerful and efficient AI techniques and emerging microelectronics technology is uniquely valuable.  Furthermore, investment in this technology needed for the grand challenges of particle physics can filter down and provide a broader impact to societal and scientific applications with less stringent requirements.  \textbf{Therefore, we advocate for continued support for: multidisciplinary research teams bringing together physicists, electrical and computer engineering, and computer scientists; development of powerful design automation tools that bring domain experts closer to the hardware and accelerates the design process; and exploration of novel and cutting-edge microelectronics implementations.}

We discuss existing work which has been recently developed for on-detector AI and the key elements of both the design and implementation, and the design tools themselves.  Though early in the exploration of these applications, the existing work highlights the needs for configurable and adaptable designs and open-source and accessible design tools to implement efficient hardware AI.  We then discuss the different classes of future applications and where investments may be needed to advance our capabilities.  We focus on novel ML techniques which are particularly important for resource-constrained real-time AI; advancing design, implementation, and verification tools which accelerate the development process; and emerging microelectronics technologies which could provide large gains in efficiency and speed.  

\clearpage
\section{Science Drivers}
 Breakthroughs in the precision and speed of sensing instrumentation enable significant advances in scientific methodologies and theories.
Thus, a common paradigm across many scientific disciplines in physics has been to increase the resolution of the sensing equipment in order to increase either the robustness or the sensitivity of the experiment itself.
This demand for increasingly higher sensitivity in experiments, along with advances in the design of state-of-the-art sensing systems, has resulted in rapidly growing big data pipelines such that transmission of acquired data for offline processing via conventional methods is no longer feasible. Data transmission is commonly much less efficient than data processing. Therefore, placing data compression, extracting waveform features and processing as close as possible to data creation while maintaining physics performance is a crucial task in modern physics experiments. 

Nearest to the sensor in particle physics experiments are often dedicated application-specific integrated circuits (ASICs) which amplify analog detector signals and digitize in analog-to-digital conversion (ADC) circuits which are then transmitted off-detector.  In cutting-edge experiments, custom ASICs are deployed due to the extreme conditions in which they must operate, in lieu of off-the-shelf electronics solutions.  For example, in energy frontier experiments such as at the Large Hadron Collider, ASICs must operate at extreme data-taking rates of 40\,MHz while maintaining low-power and operating in a very high radiation environment.  In next generation neutrino detectors,  quantum computing and sensing applications, and cosmology and direct detection dark matter experiments, ASICs processing detector signals must operate in cryogenic environments and also are required to be very low power.  Furthermore, similar technologies are needed in modern nuclear physics experiments such as the Electron-Ion Collider (EIC) and X-ray discovery science experiments at light sources. As the number of sensing channels continues to grow, driving experiments to explore nature at finer spatial and temporal scales, the amount of on-detector processing required will continue to increase.  


Artificial intelligence (AI), and more specifically machine learning (ML), has recently been demonstrated to be a powerful tool for data compression, waveform processing~\cite{Miryala_2022}, and analysis in physics~\cite{Albertsson:2018maf,Radovic:2018dip,Bourilkov:2019yoi,Carleo:2019ptp} and many other domains.
While progress has been made towards generic real-time processing through inference including boosted decision trees and neural networks (NNs) using FPGAs (Field Programmable Gate Arrays) in off-detector electronics~\cite{hls4ml,CMSP2L1T}, ML methods are not commonly used to address the significant bottleneck in the transport of data from front-end ASICs to back-end
FPGAs. 
\textbf{Due to historical requirements for processing data in extreme environments in high energy physics experiments, the advancement of AI in near-detector electronics is a natural evolution of detector readout in the field and can be a driver for progress in other scientific domains.
Embedding ML as close as possible to the data source has a number of potential benefits}:
\begin{itemize}
    \item ML algorithms can enable powerful and efficient non-linear data reduction or feature extraction techniques, beyond simple summing and thresholding, which better preserves the physics content that would otherwise be lost;  
    \item This could in turn reduce the complexity of down stream processing systems which would then have to aggregate less overall information all the way to offline computing; 
    \item This enables real-time data filtering and triggering like at the LHC and the EIC which would otherwise not be possible or be much less efficient; or in the case of cryogenic systems, creates less data bandwidth from cold to warm electronics and thus reduce the system complexity;
    \item Furthermore, intelligent processing as close as possible to the source will enable faster feedback loops. For example, in continuous learning applications, if the data is part of a control or operations loop where feedback is needed such as in quantum information systems or particle accelerators. 
\end{itemize}

In the increasingly complex big data experiments and accelerators which are among the most complex scientific instruments ever made, real-time on-detector AI will be necessary to keep up with more sensitive detectors and expand their capabilities to accelerate scientific discovery.


\section{Community Needs}
\label{sec:needs}
There has been a long interest in deploying AI techniques in readout electronics in high energy physics.  Explorations in the early 1990s resulted in a number of early implementations~\cite{CERNCourierVolume32:1732048} for real-time triggering and machine controls at Fermilab, SLAC, DESY, CERN, and BNL.  There are a number of reasons for a recent renewal of these efforts that will enable even more powerful AI on-detector implementations: 
\begin{itemize}
    \item \textbf{broader necessity}: as Moore's Law stalls, we cannot solely rely purely on advances offline, data center, or ``cloud'' computing; higher energy physics is no longer the only customer for such ``edge'' or ``Internet-of-Things'' AI applications - both in scientific and industrial applications; this has led to technological progress on which we can capitalize
    \item \textbf{advances in hardware}: in the past 30 years, hardware technology nodes and manufacturing reliability has advanced energy efficiency and speed with modern chips at the sub-10\,nm technology node.
    \item \textbf{advances in codesign tools}: electronic design automation tools and abstraction has greatly progressed to make designing custom electronics much more expedient; this includes advances such as High Level Synthesis (HLS) tools
    \item \textbf{advances in machine learning methods}: advances in machine learning theory itself has made for much more efficient and powerful implementations of neural network architectures including concepts such as quantization, compression, and neural architecture search.  
\end{itemize}

While the emergence of AI is growing rapidly across a  number of domain areas, \textbf{the requirements of particle physics for speed, throughput, fidelity, interpretability, and reliability in extreme environments requires advancing state-of-the-art technology far beyond industrial and commercial applications}.  Therefore, dedicated investment in powerful and efficient AI techniques and emerging microelectronics technology is uniquely important in particle physics.  Furthermore, investment in this technology needed for the grand challenges of physics can provide a broader impact to less stringent applications in science and industry.  

\begin{figure}[tbh!]
    \centering
    \includegraphics[width=0.7\textwidth]{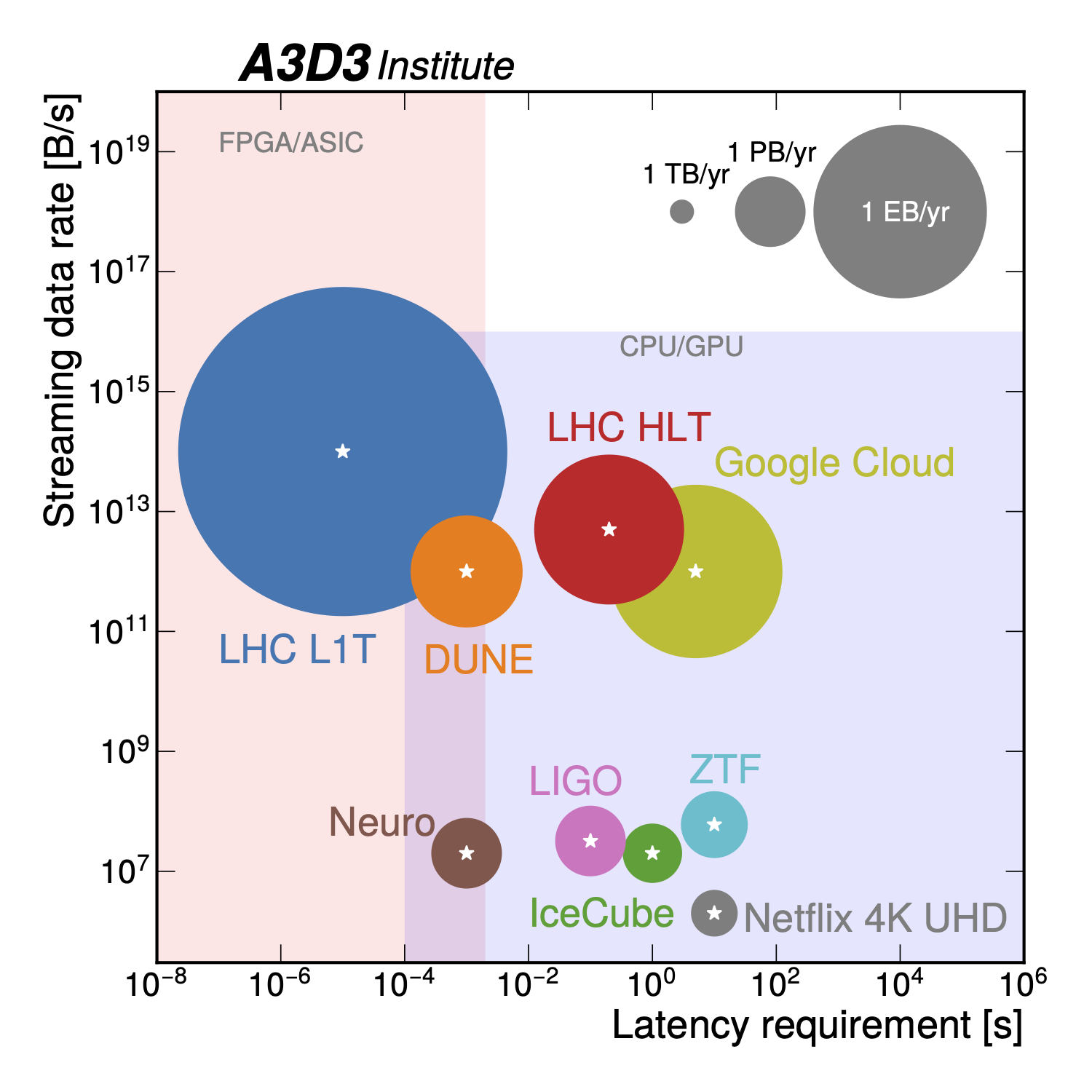}
    \caption{Latency, throughput, and estimated dataset sizes for a number of scientific and industry real-time AI applications, provided by the A3D3 NSF institute~\cite{a3d3}}
    \label{fig:a3d3}
\end{figure}

To illustrate this for specific examples, Fig.~\ref{fig:a3d3} shows the latency and throughput requirements and the anticipated dataset sizes for a number of particle physics experiments as well as industry applications. The particle physics real-time AI  applications have much more challenging throughput and latency requirements.  Solving these challenges will necessitate the development of novel design and hardware solutions.  We have also seen that open-source tools developed for science can have an impact on industrial and mainstream applications.  The {\tt hls4ml} workflow, developed for science and described in Sec.~\ref{sec:existing}, has been shown to be a powerful solution in the MLPerf\textsuperscript{TM} Tiny AI benchmarks~\cite{mlperf}.  These benchmarks are focused on edge and IoT applications such as voice assistants, wearable technology, and Industry 4.0.  

Therefore, we believe that investment in particle physics for on-detector AI is necessary and unique and would potentially enable advances that other scientific domains and society could benefit from.  In particular, \textbf{we advocate for continued support for: multidisciplinary research teams bringing together physicists, electrical and computer engineering, and computer scientists; development of powerful design automation tools that bring domain experts closer to the hardware and accelerates the design process; and exploration of novel and cutting-edge microelectronics implementations.} In Section~\ref{sec:existing}, we will discuss existing work related to deploying AI techniques for ASICs in particle physics which introduces modern applications and workflows.  In Section~\ref{sec:apps}, we will discuss applications and new directions for research in both design tools and hardware exploration.  

\section{Existing Work}
\label{sec:existing}
We discuss some recent, modern efforts in deploying AI techniques in readout electronics and the tools being developed.  It is still early in this exploration, however, and we will continue to see advancements in the near future.  

\subsection*{Reconfigurable data compression for HL-LHC}

At the CERN Large Hadron Collider (LHC) and its high luminosity upgrade (HL-LHC), extreme collision rates present extreme challenges for data processing and transmission at multiple stages in detector readout and trigger systems. As the initial stage in the data chain, the on-detector (front-end) electronics that read out detector sensors must operate with low latency and low power dissipation in a high radiation environment, necessitating the use of application-specific integrated circuits (ASICs).
In order to mitigate the initial bottleneck of moving data from front-end ASICs to off-detector (back-end) systems based on field-programmable gate arrays (FPGAs), front-end ASICs must provide edge computing resources to efficiently use limited bandwidth through real-time processing and identification of interesting data. Front-end data compression algorithms have historically relied on zero-suppression, threshold-based selection, sorting or summing of data.  Recently, for the CMS (Compact Muon Solenoid) endcap calorimeter upgrade, an ASIC for data compression using a reconfigurable autoencoder was developed; it is described in much more detail in Ref.~\cite{DiGuglielmo:2021ide}.    

The ASIC ML algorithm uses a traditional autoencoder algorithm which creates an information bottleneck that compresses sensor data.  The encoder part of the autoencoder, with a neural network architecture consisting of a convolutional layer and a dense layer, was implemented in the ECON-T data concentrator ASIC.  One key feature of the design are that the parameters (weights and biases) of the encoder are \textit{reconfigurable}.  Through reconfiguration, we will be able to: 
\begin{itemize}
    \item \textit{enable} more computationally complex compression algorithms, which could improve overall physics performance or allow more flexible algorithms;
    \item \textit{customize} the compression algorithm of each sensor based on their location within the detector;
    \item \textit{adapt} the compression algorithm for changing detector and collider conditions (for example, if the detector loses a channel or has a noisy channel it can be accounted for or if the collider has more pileup than expected, the algorithm can be adjusted to deal with new or unexpected conditions without catastrophic failure).
\end{itemize}

Simulations of the chip meets latency, throughput, area, and power requirements while being radiation tolerant.  The network parameters are triplicated using triple modular redundancy in order to mitigate radiation effects from single event upsets.  The design has a latency of 50\,ns with an initiation interval of 25\,ns.  The design has an area of 2.88\,mm$^2$ and consumes 48\,mW of power which meets area and power requirements which are 4\,mm$^2$ and 100\,mW respectively.  Physics performance studies are still under study but initial studies show that it does a good job of reconstruction energy patterns in the sensor using the Energy Mover's Distance metric~\cite{Komiske:2019fks}.  The ECON-T chip has been fabricated and is currently under test.    

\subsection*{Software-hardware codesign with hls4ml}

\begin{figure}
\centering
\includegraphics[width=0.95\textwidth]{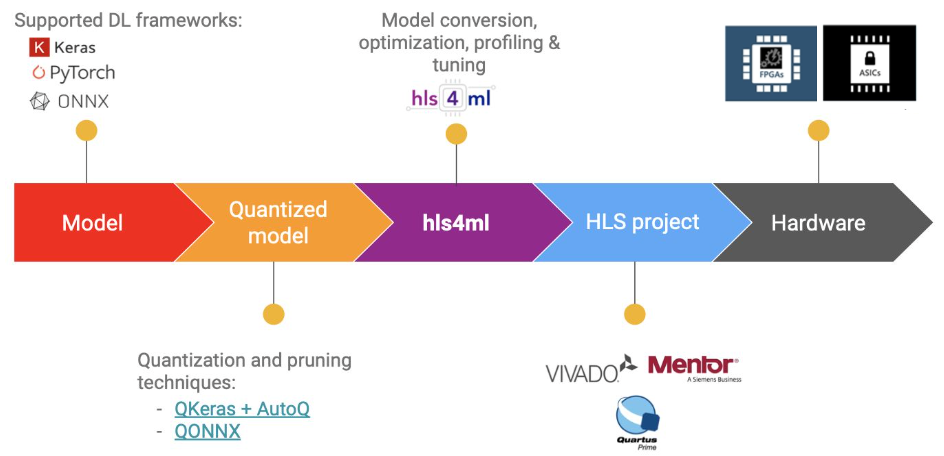}
\caption{A typical workflow to translate an ML model into an FPGA or ASIC implementation using {\tt hls4ml}.
The red boxes (left) describe the model training and compression steps performed within conventional ML software frameworks.
The configuration and conversion steps are shown in the blue boxes (center).}
\label{fig:flow}
\end{figure}

To accelerate the implementation of ML algorithms in hardware, we need tools that will translate the algorithms into hardware primitives used in ASIC design. To that end the ECON-T encoder algorithm used the {\tt hls4ml} tool~\cite{Duarte:2018ite,hls4ml_github}, a codesign framework that translates trained NNs, specified by the model's architecture, weights, and biases, into the specification of a hardware accelerator that can be synthesized with HLS tools.
Figure~\ref{fig:flow} shows the schematic of a typical workflow.
The first part of the workflow, performed with tools like \textsc{(Q)Keras}~\cite{Coelho:2020zfu,chollet2015keras} and \textsc{PyTorch}~\cite{paszke2019pytorch} or \textsc{Brevitas}~\cite{brevitas}, involves training and compression steps before converging on a final model.
Efficient techniques for compressing the model include reducing the precision of the computations in ML algorithms either through post-training quantization or quantization-aware training as well as model pruning -- removing unimportant synapses or neurons in a neural network.  

The conversion step is performed with {\tt hls4ml}, which translates a model into an HLS (High Level Synthesis) project that can subsequently be synthesized and implemented on an FPGA (Field Programmable Gate Array) or ASIC.  Generally, HLS tools translate algorithms from higher level programming languages such as C or Python into a lower-level hardware description languague (HDL).
In the case, for example, of the ECON-T ASIC autoencoder, the {\tt hls4ml} tool used the Mentor Siemans Catapult HLS tool as a backend to implement the neural network for the ASIC.
At this stage, a variety of configurable parameters can be tuned, including the precision and level of parallelization layer-by-layer.
A suite of \textit{optimizers} then modify the network graph to target a more lightweight, faster inference.
Beyond this ASIC applications, {\tt hls4ml} has been used for a range of use-cases from particle accelerator control~\cite{John:2020sak}, to LHC L1 trigger~\cite{CERN-LHCC-2020-004}, incremental model training~\cite{abdulqader2021enabling}, and on-detector data compression~\cite{DiGuglielmo:2021ide}.
It is a unique aspect of {\tt hls4ml} to support multiple vendor backends (e.g. Xilinx~\cite{vivadohls}, Intel~\cite{quartus2020}, and Mentor~\cite{catapulthls2020}) with possible expansion to others.

\subsection*{Neural Networks for Waveform processing}
\begin{figure}[htbp]
\centering
\includegraphics[width=0.90\columnwidth]{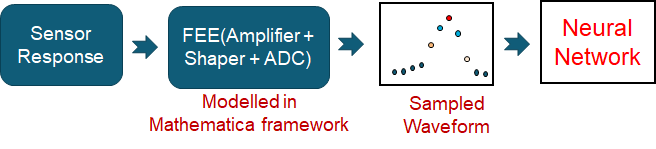}
\caption{Model of real-time feature extraction using an ANN. 
}
\label{fig:ANN_processing}
\end{figure}

\begin{figure}[htbp]
\centering
\includegraphics[width=0.9\columnwidth]{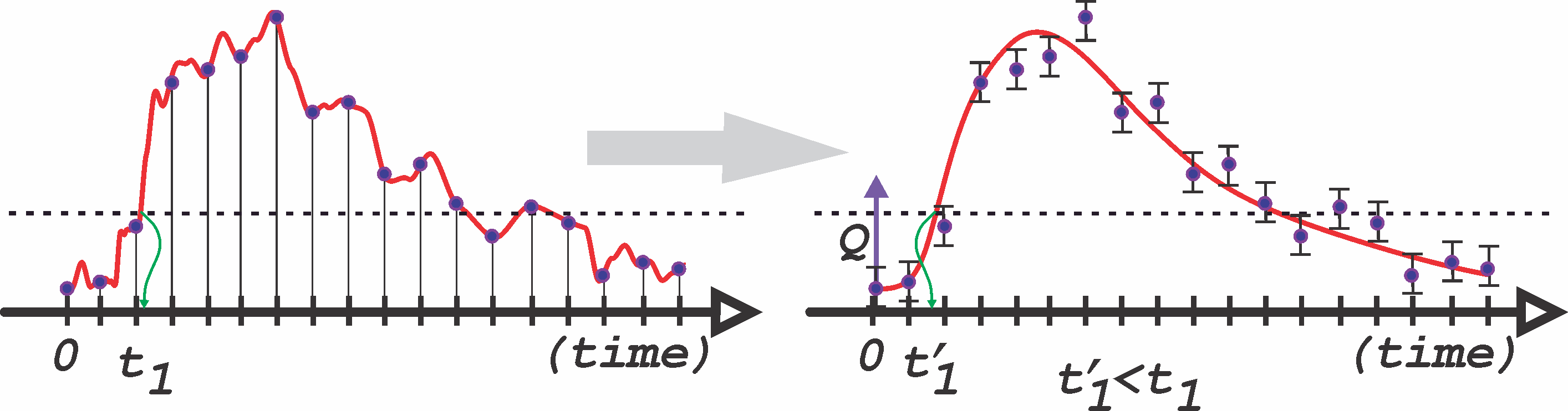}
\caption{Sampling (left) and fitting (right) of a noisy pulse waveform.}
\label{fig:ADC_sample}
\end{figure}

Neural network implementations in front end detector electronics can be a powerful tool to overcome the limits of traditional readout approaches. These neural networks can be trained and used as a computational block along with the signal processing path to provide detailed information about the radiation signals, i.e., high quality data. Figure~\ref{fig:ANN_processing} shows a block diagram of a neural network training setup with high-level modeling of all the signal processing blocks. The neural network is initially trained with simulated data, and training can be further improved with real data once the actual experiment is underway. A waveform is the product of the continuously-operating front-end processing chain. Such waveforms contain characteristic features of the radiation signal, such as {\em amplitude} over {\em time}. The neural network is trained to extract the most common radiation signal parameters: {\em energy}, {\em time of arrival}, {\em charge sharing}, {\em pile up}, etc. In the rest of this section, we describe some use cases of neural networks in detectors currently being developed at BNL.
The impinging radiation releases charge packets that after analog processing take the shape of pulses. However, implementing analog processing in sub 100~nm process nodes is becoming increasingly challenging. At the same time, high resolution analog-to-digital conversion is difficult to achieve due to the area and power constraints of densely-arranged channels of pixel detectors. Waveform sampling of asynchronous signals cannot guarantee proper sampling of the waveform peak, hence the samples need to be interpolated to reconstruct the key features of the waveform, such as the peak. However, the peak of the waveform is not necessarily proportional to the deposited energy, therefore the interpolation has to account for the pulse shape to accurately estimate the energy deposited by the impinging radiation. Such interpolation can be achieved through deconvolution, which is effectively a process of  ``fitting'' of a predefined curve to the available samples of the waveforms as visualized in Fig.~\ref{fig:ADC_sample}. The accuracy of the deconvolution, which is limited in the presence of noise, can be improved using non-linear corrections~\cite{Sajedi2013ANN}. This task is thus suitable for an ANN. The ANN approach is expected to exceed the precision of conventional fitting, which is limited by the low resolution conversion of individual samples. 

\begin{figure*}[tb]
\centering
\includegraphics[scale=0.2]{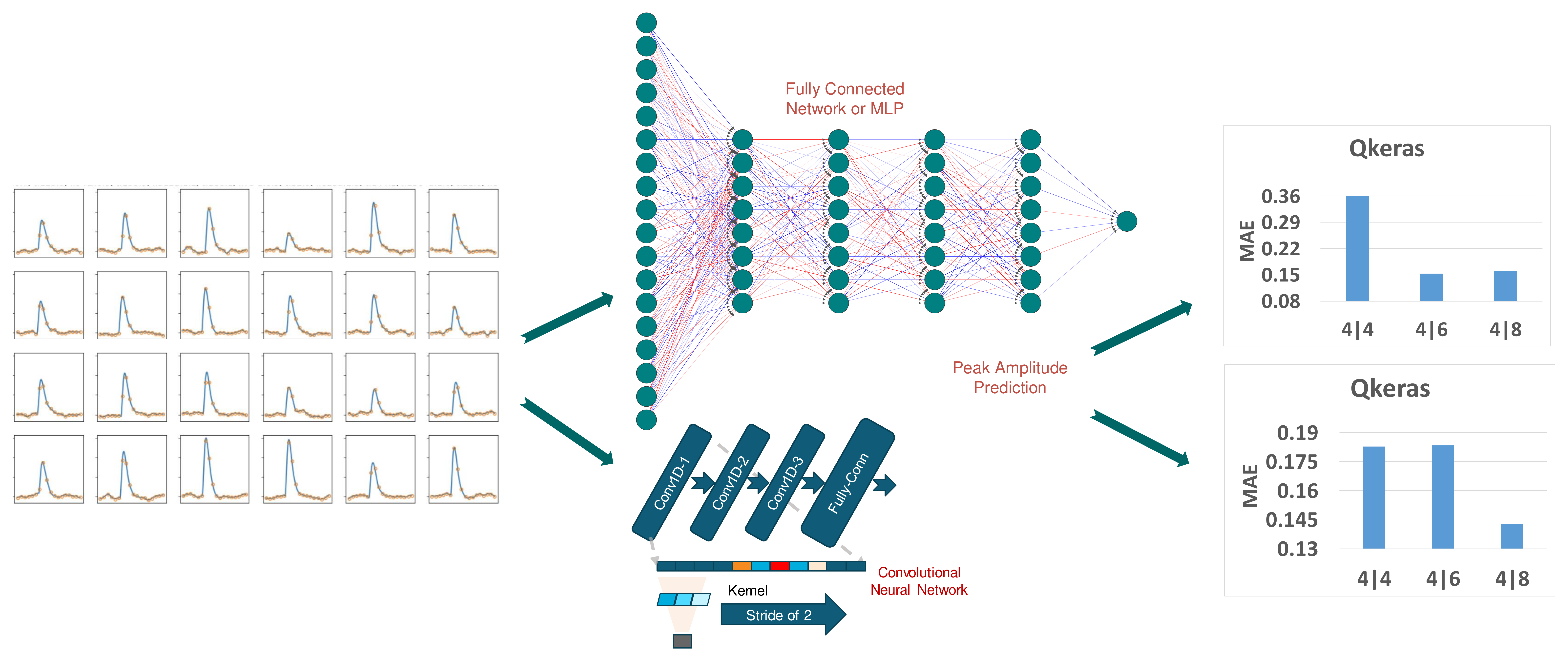}
\caption{Proposed ANN-based  methodology and Mean Absolute Error (MAE) estimation~\cite{Miryala_2022}. In the MAE plots, ($m | n$) represents $m$ bit ($n$ bit) precision for first (last) two hidden layers.}
\label{fig:ann_methodology}
\end{figure*}

The goal is to estimate the energy deposited by the impinging radiation from single waveform snippets with high-accuracy inference while using the lowest sampling rate and lowest conversion resolution possible. To that end, we study the use of ANN-based approaches for extracting information from signal pulses. Artificial neural networks such as multi-layer perceptrons (MLPs), convolutional neural networks (CNNs) can be trained for feature extraction from the pulses. We calculate and mathematically generate the pulses, as theoretically produced by a silicon sensor, with the expected information of the deposited energy, and use them as the training data. We add noise that represents the actual system to the pulses. We also include non-idealities in the transfer function of the pulse forming filter.

In order to minimize the ADC sampling rate, we typically use three or four samples on both the rising and falling edges of the pulses. The waveforms are randomly partitioned into training, validation, and test subsets. We utilize 90\% of the data for training and validation and the remaining 10\% for testing. Two well-accepted ANN algorithms are evaluated (see Fig.~\ref{fig:ann_methodology}), namely MLPs and CNNs. Preliminary studies demonstrated that the two ANN algorithms investigated are effective in feature extraction using pulse data sets produced in simulations of the combined responses of a silicon sensor and second order shaping filter. 

\section{Applications, Design, and Technology}
\label{sec:apps}
\subsection{System-level use-cases}

There are a few potential possibilities for integrating AI into the on-detector electronics.  In this white paper, we consider only applications on the detector and leave AI applications in off-detector electronics and computing to other studies and papers. Here, we discuss in some detail two main classes where we consider deploying AI: sensor-integrated AI and on-detector data concentration.  There are scenarios that could be considered in-between these two classes or between near- and off-detector electronics, but we consider these categories for the sake of classification and discussion.

\paragraph{Sensor-integrated AI}

Readout electronics integrated with the sensor is often required for highly granular applications.  Examples include pixel detectors with readout electronics bump-bonded to the sensor~\cite{Dinardo_2015}, MAPS (Monolithic Active Pixel Sensors)~\cite{TURCHETTA2006139}, and high-granularity calorimeter sensors in CMS~\cite{CERN-LHCC-2017-023}.  The role of readout electronics in these systems are to reduce data at very low latencies and rely only on local information.  Typically, these readout ASICs are performing analog-to-digital-conversion and reduction of digital data to manageable rates for transmission off the sensor.  As channel counts per sensor continue to grow, performing feature extraction or data compression on sensor has become increasingly important.  

AI algorithms could operate on the signals after digitization and should be highly integrated with which information is made available.  Alternatively, there is also the potential to perform AI before digitization - on the analog signals themselves.  This would require more speculative technologies such as analog AI circuits and consider neuromorphic types of algorithms, but could potentially be much more power efficient.  We will discuss those techniques more in Sec.~\ref{sec:future}.

\paragraph{On-detector data concentration}

Often in on-detector data concentration, the data has been digitized and is being aggregated across multiple channels or sensors.  An example that we have discussed above is the ECON-T.  Requirements on size, weight, area, and power are typically less stringent than the sensor-integrated AI.


\subsection{Efficient machine learning training and implementation}

The field of machine learning is vast and rapidly evolving. 
Because of this, in this subsection, we will call attention to a few particularly important techniques that are needed for ASIC implementations of machine learning algorithms. More specifically, we will focus on the following two techniques: (1) how to train energy efficient neural networks and, (2) the importance of domain adaption to support ASIC reconfigurability.

\subsubsection*{Reduced precision and compression}

\textit{For a more detailed review of these ML techniques, please see Ref.~\cite{Deiana:2021niw} and references therein written by Amir Gholami, Zhen Dong, and Sehoon Kim -- what follows here is a very reduced summary.}

Executing machine learning algorithms in reduced precision, i.e. quantization~\cite{asanovic1991experimental, hubara2016binarized, rastegari2016xnor, zhou2017incremental, zhou2016dorefa, jacob2018quantization, zhang_2018_lqnets, dong2019hawq, dong2019hawqv2,yao2020hawqv3,krishnamoorthi2018whitepaper,naumov2018periodic, elthakeb2020gradient,banner2019post,wang2018haq,hubara2020improving}, and through compression, i.e. sparsity/pruning~\cite{hassibi1993optimal,han2016eie,luo2017thinet,liu2018rethinking,dong2017learning,zeng2018mlprune,lecun1990optimal,han2015learning, molchanov2016pruning, li2016pruning, mao2017exploring, yang2017designing,wang2020differentiable,tung2018clip,guerra2020automatic,hacene2018quantized,kwon2020structured,wang2019eigendamage,zhu2017prune}, enables us to optimize the NN significantly. 
In quantization, low precision is used to represent the weights and
activation, whereas in pruning the connections are completely removed.
Quantization comes in two main classes: (a) post-training quantization where training is performed traditionally in 32-bit floating point operations and then after training, the ranges of weights and activations are reduced adiabatically without loss of performance; or (b)  quantization-aware training where during training time, the losses are computed including reduced precision.  
Quantization-aware training requires more care, but it has been shown to require less implementation resources than post-training quantization~\cite{qkeras}.  
The common goal of both quantization and pruning is to dispense a given budget of resources for the maximal benefit of the entire data reduction path.
Optimizations applied on the data reduction path will help to expose a maximal set of network components to pruning and quantization while avoiding irreconcilable discrepancies between the theoretical and practical performance of the NNs. 

While there has been work in quantization/pruning, many existing approaches focus on industry applications and involve ad-hoc methods that do not generalize to new models/tasks, especially for cases with high degrees of compression (i.e., ultra low bit precision for quantization and high sparsity for pruning).
This problem is being addressed more generically for quantization in recent work
on HAWQ and Q-BERT~\cite{dong2019hawq,dong2019hawqv2,yao2020hawqv3,shen2020q,kim2021bert}, where they showed a theoretical connection to Hessian spectrum.
In particular, layers with smaller Hessian spectrum can be quantized to lower precision, and vice versa.
Similar connections were also made for pruning~\cite{lecun1990optimal,hassibi1993optimal,zeng2018mlprune,wang2019eigendamage}. 

Furthermore, a largely missing piece is hardware-aware codesign of quantization and pruning.
While there has been some work in the literature~\cite{wang2018haq,wang2020hat}, this 
work typically uses simulated hardware metrics and/or relies on methods based on reinforcement learning (RL) to find the compression scheme. 
Another important challenge is that there is very little theoretical work to study the combination of these methods, and how much performance drop is expected given a NN model and a compression scheme.

\subsubsection*{Domain adapation and transfer learning}

ASICs are hardened integrated circuits designed for optimization of specific tasks.  Often, rule-based algorithms in ASICs are simplistic and have only a few parameters, if any, which can be tuned.  In the case of ML, algorithms are characterized by their parameters (weights and biases).  Making ML parameters reconfigurable in an ASIC could be very expensive.  However, totally hard-coding the weights can leave the algorithm unable to adapt to changing detector conditions.  

Partial configurability also remains an option and is sometimes referred to as fine-tuning or transfer learning~\cite{Pan2010ASO,Tan2018ASO}.  For example, in large computer vision networks, when adapting a model for different tasks, the early layers in an ML model are fixed while only the parameters in the latter layers are retrained.  
Hybrid techniques such as these are important to explore in ASIC ML algorithms because any weights that can be fixed will greatly reduce area and energy resources.  

\subsection{AI co-design tools and methodology}

\paragraph{Neural network translation and representation lowering}

In Sec.~\ref{sec:existing}, we discussed the {\tt hls4ml} codesign workflow which starts with common ML software frameworks such as (Q)Keras, TensorFlow, and PyTorch and translates NN models into a hardware description.  While that workflow has been demonstrated for a number of different applications, it, or other similar workflows, can be further expanded.  There are a number of interesting directions to advance available tools:
\begin{itemize}
    \item There are a number of other HLS toolflows which support ASICs including Cadence Stratus HLS and the existing Catapult HLS tool flow should also be expanded in its capabilities.  Furthermore there are other potential high-level languages which can be used for hardware implementations such as BlueSpec~\cite{1459818} and Chisel~\cite{10.1145/2228360.2228584}.  These various  tools enable expressing neural network operations at a higher level.  There is also recent work in representing networks more natively for hardware such as QONNX~\cite{qonnx}, a specialized neural network interchange format with flexibility for heterogeneous quantization. Other compiler/transpilation tools such as LLVM/MLIR~\cite{lattner2020mlir} offer promise as well and tools like SODA are exploring such ecosystems~\cite{10.1145/3400302.3415781}.  
    \item The above workflows thus far only focus on standard CMOS technologies.  For future technologies such as analog AI, the tool flows are much less mature and require more effort to interface and make available for non-experts.  Such technologies are discussed more in Sec.~\ref{sec:future}.
\end{itemize}

The proliferation of AI techniques in custom detector systems requires an ecosystem of easy-to-use tools which enable rapid development between domain scientists, computer scientists, and electrical and computer engineers.  Furthermore AI methods can themselves also improve and enable more efficient hardware design.  In an era of rapidly developing research in computing micro-architectures, it is valuable to maintain strong collaborations with developers of these tools which are often outside of high energy physics.  

\paragraph{Integration, system-on-chip, and verification}

\begin{figure}[tbh!]
    \centering
    \includegraphics[width=0.60\textwidth]{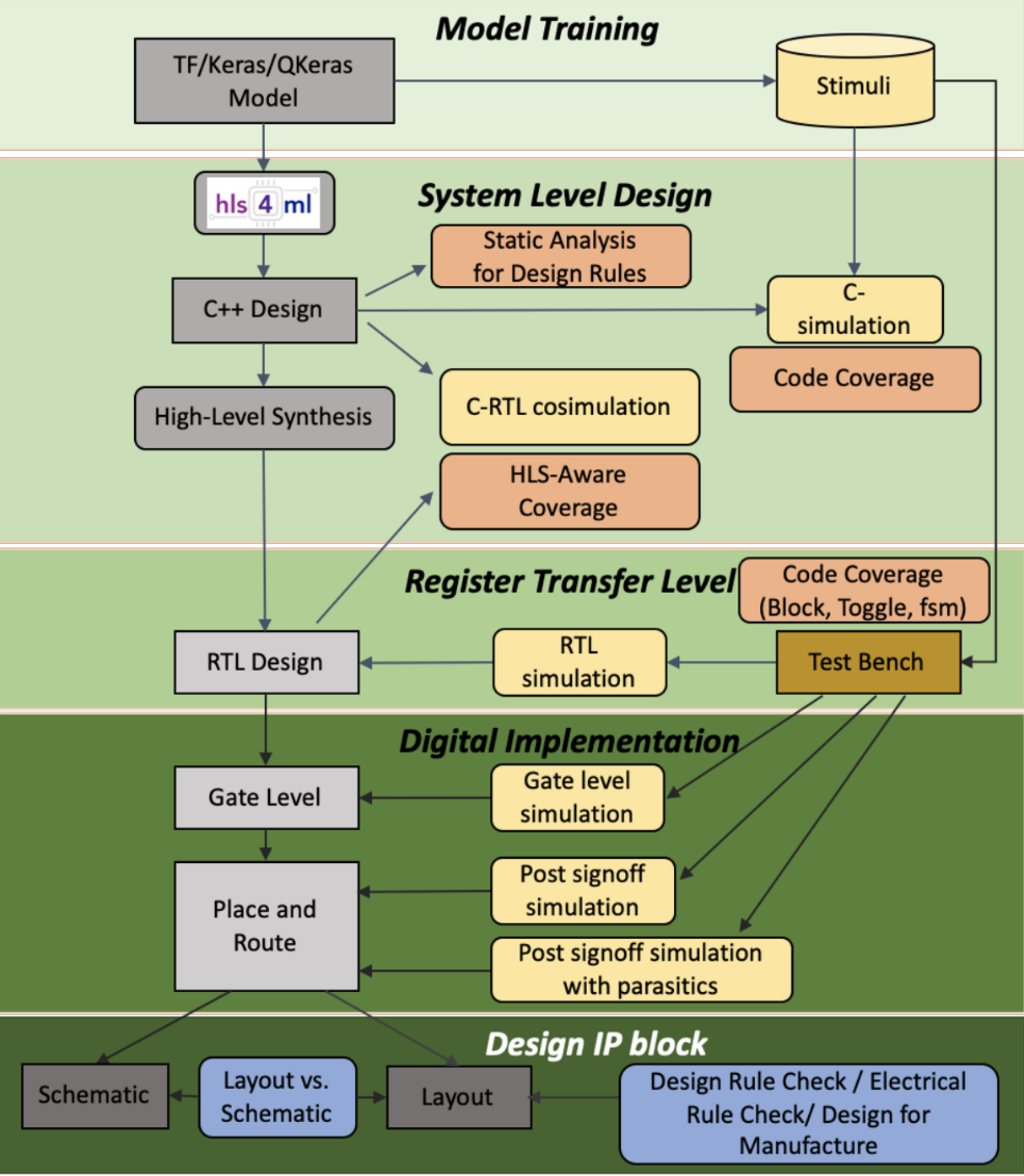}
    \caption{The full implementation chain of the ECON-T autoencoder
    \label{fig:validation}}
\end{figure}

The full integration and validation workflow for the ECON-T autoencoder is shown in Fig.~\ref{fig:validation}.  Similar to continued investment and research in codesign workflows discussed above, it is also important to continue developing design automation tools for implementation and validation.  Figure~\ref{fig:validation} highlights the need for extensive validation and design tools to accelerate the development cycle while maintaining reliable verification techniques.  These tools are often ad hoc or integrated into proprietary software tools from EDA vendors.  

Furthermore, system-level integration tools can greatly accelerate development.  For example, system-on-chip (SoC) platforms offer a powerful flexibility for on-sensor algorithm orchestration and data flow. Projects like the Embedded Scalable Platform (ESP) is an open-source tools which can greatly accelerate the development of SoC in a flexible system-wide methodology.  
Another example of a project exploring SoC is the sparkPix \cite{rota2021iWoRID} detector that includes the integration of sensor, ADCs, processing logic in the analog domain and data flow management at 1-MHz frame rate. This platform is being extended to include digital processing and compression algorithms using non-AI techniques that could complement the AI based developments. 


\subsection{Emerging microelectronics technologies}
\label{sec:future}
\subsubsection*{Advance Geometry nodes}
With rapidly growing machine learning applications comes the acute need for their efficient hardware implementations. Most of the efforts are focused on digital CMOS technology, such as implementations based on general-purpose TPUs/GPUs, FPGAs, and more specialized ML hardware accelerators.  The steady improvements in such hardware platforms' performance and energy efficiency over the past decade are attributed to the use of very advanced, sub-10-nm CMOS processes and holistic optimization of circuits, architectures, and algorithms. 
It includes, for example, taking advantage of aggressive voltage supply scaling \cite{Moons2017}, very deep pipelines and extensive data reuse in architectures \cite{Chen2017}, and lowering the precision of weights and activations of the algorithms \cite{Simons2019}.  As a result, very compact state-of-the-art neural networks, such as MobileNet based on 3.4M parameters and 300M multiply-and-add operations per inference \cite{Sandler2018}, can now be fitted entirely on a single chip. However, on all these fronts, advances are saturating and cannot rely on the faltering Moore's law. 
Advanced geometry nodes such as 28 nm and below are currently being investigated for Phase III upgrades of HL LHC. A community driven effort for modelling radiation effects led by INFN and CERN is currently underway. 
Similarly, for the fully depleted 22 nm FDSOI process, Fermilab is developing cryogenic models at 4K with EPFL and Synopsys. The back gate control available in 22 FDX allows digital operation at ultra low supply voltages of 400 mV and below.

\subsubsection*{Beyond CMOS}

\textit{For a more detailed review of beyond CMOS technologies, please see Ref.~\cite{Deiana:2021niw} and references therein written by Dmitri Strukov -- what follows here is a very reduced summary.}


The opportunities for building more efficient hardware may come from biological neural networks. Indeed, it is believed that the human brain, with its $>$1000$\times$ more synapses than the weights in the largest transformer network, is extremely energy efficient~\cite{Hasler2013}, which serves as a general motivation for developing neuromorphic hardware~\cite{Mead1990}. There is a long history of CMOS neuromorphic circuits~\cite{Indiveri2011}. However, unleashing the full potential of neuromorphic computing might require novel, beyond-CMOS device and circuit technologies \cite{Berggren2020} that allow for more efficient implementations of various functionalities of biological neural systems. 

The most prominent emerging technology proposals, including those based on emerging dense analog memory device circuits, are grouped according to the targeted low-level neuromorphic functionality - see, e.g. reviews in \cite{Burr2017, Bavandpour2018, Yang2013NatureNano, Yu2018IEEE} and original work utilizing volatile \cite{Sheridan2017, Cai2019NatElec, Chu2014Neuro, Yeon2020, Ohno2011, Wang2017NatMat, Pickett2013, Wang2018NatElec, Zhang2018Small, Lashkare2018, Adda2018} and nonvolatile \cite{Alibart2012, Adam2017,Govoreanu2013, Prezioso2015, Prezioso2016, Prezioso2018, MerrikhBayat2018, Lin2020NatElec, Hu2018AdvMat, Yao2020Nature, Liu2020ISSCC, Kim2019XBAR, Cai2020NatElec, Mahmoodi2019IEDM, Mahmoodi2019NatComm, Li2016IEDM, Wang2018NatElec, Pedretti2017}  memristors, phase change memories (PCM) \cite{Burr2015, Tuma2016, Ambrogio2018, Karunaratne2020, Joshi2020, Kuzum2011, Rios2019}, and nonvolatile NOR \cite{Guo2017CICC, Guo2017IEDM, MerrikhBayat2015, Mahmoodi2019NatComm}, and NAND \cite{Bavandpour2019NAND, Bavandpour2020, Lee2019NAND}, and organic volatile \cite{Fuller2019} floating gate memories, as well as multiferroic and spintronic \cite{Grollier2020, Ostwal2018, Sengupta2016, Romera2018, Ni2018IEDM}, photonic \cite{Shasti2021, Goi2020, Rios2019, Lin2019Sciece, Hamerly2019PhysRevX, Hamley2019, Shen2017NatPhot, Tait2016, Feldmann2019, Buckley2017, Bruiner2013, Vandoorne2014}, and superconductor \cite{Segall2017, Buckley2017, Rowlands2021} circuits. 
Many emerging devices and circuit technologies are currently being explored for neuromorphic hardware implementations. 
Neuromorphic inference accelerators utilizing analog in-memory computing based on floating gate memories are perhaps the closest to widespread adoption, given the maturity of such technology, the practicality of its applications, and competitive performance with almost 1000x improvement in power as compared to conventional (digital CMOS) circuit implementations ~\cite{Hasler2013}. The radiation hardness of these techniques and their applicability for robust performance in extreme environments is yet to be evaluated.








\bibliographystyle{JHEP}
\bibliography{refs}






\end{document}